\documentclass[journal=cdgefu,manuscript=article]{achemso}

\usepackage[version=3]{mhchem}
\usepackage{bm}
\usepackage{amsmath}
\usepackage{amssymb}

\author{Adriana Val\'{e}rio}
\affiliation{Institute of Physics, University of S{\~{a}}o Paulo, S{\~{a}}o Paulo, SP, Brazil}

\author{S{\'{e}}rgio L. Morelh{\~{a}}o}
\email{morelhao@if.usp.br}
\affiliation{Institute of Physics, University of S{\~{a}}o Paulo, S{\~{a}}o Paulo, SP, Brazil}

\title{Usage of Scherrer's formula in X-ray diffraction analysis of size distribution in systems of monocrystalline nanoparticles}

\begin{document}

\begin{abstract}
In the supporting information file of the article Controlled Formation and Growth Kinetics of
Phase-Pure, Crystalline BiFeO3 Nanoparticles (Crystal Growth \& Design 2019), there is a description on how to use Scherrer equation for in situ X-ray diffraction analysis of crystallization processes investigated in the article. That description led to a necessary revaluation on the current understanding of the usage of Scherrer equation for analyzing size distributions, as discussed in this work.
\end{abstract}

\section{INTRODUCTION}

In X-ray diffraction analysis of powder samples, diffraction peak width $\beta_s$ provides estimation of crystallite size (size of small crystals, grains, or particles diffracting in powder samples) when deconvolved from instrumental broadening $\beta_{\rm inst}$ according
to $\beta_s = \left(\beta^2_{\rm exp}-\beta^2_{\rm inst}\right)^{1/2}$ where $\beta_{\rm exp}$ is the actual measurement of the peak full width at half maximum (fwhm). Scherrer equation (SE)\cite{ps18} leads to the crystallite size \begin{equation}\label{eq:SE}
L_s = \kappa\, \lambda/\beta_s\cos(\theta_B)
\end{equation}
where $\kappa$ stands for a geometrical factor that depends on crystallite apparent radius of gyration\cite{sm16} from the perspective of reflections with Bragg angle $\theta_B$ for X-rays of wavelength $\lambda$. For instance, crystallites with cubic shape have $\kappa\simeq0.92$, while spherical crystallites have $\kappa\simeq1.18$. However, real systems of crystalline nanoparticles are, in the vast majority of cases, formed by crystallites of different sizes. What does the $L_s$ value obtained from SE stand for in these cases? It has been understood that, when measuring diffraction peak widths in powder samples with particle size distribution (PSD), the  crystallite size $L_s$ obtained from SE, Eq.~(\ref{eq:SE}), corresponds to the average value of the volume-weighted PSD.\cite{bw90,ck98,ac05,mb09} In this work, a simple line profile simulation of X-ray diffraction peaks in samples with PSD leads to a different understanding. It shows that SE provides $L_s$ as the median value of the weighted PSD by crystallite sizes to the power of four. Because this weighting by the crystallite size to the power of four, larger particles have tremendous contribution to the median value. As dynamical diffraction effects (absorption and re-scattering) or, as formerly called, primary extinction, can significant diminish intensity contributions of crystallites with submicron dimensions, a method to account for primary extinction corrections when analyzing PSD in systems of monocrystalline particles is also discussed.

\section{THEORETICAL APPROACH}

\subsection{Kinematical diffraction}

According to the kinematical theory of x-ray diffraction,\cite{bw90,sm16} the integrated intensity $P_c=\int I_c(2\theta)d2\theta$ of each diffraction peak $I_c(2\theta)$ from a single crystallite (small crystal or particle diffracting in powder samples) as a function of the scattering angle $2\theta$ is proportional to the crystallite volume $V_c$, that is $P_c\propto V_c$. On the other hand, while the peak area increase with $V_c$, the peak width (fwhm) $\beta$ gets narrower inverselly with crystallite size $L$, that is $\beta \propto 1/L$. To account for the fact that actual powder samples are formed by crystallites with different sizes, the number of diffracting crystallites with size between $L$ and $L+dL$ is given by $n(L)dL$ where $n(L)$ is the PSD function so that
\begin{equation}\label{eq:N}
    N=\int n(L)dL
\end{equation}
is the total number of diffracting particles in the sample.

By using normalized line profile functions, such as a Lorentzian function
\begin{equation}\label{eq:lpf}
    \mathcal{L}(2\theta) = \beta^2/\left [ 4(2\theta - 2\theta_B)^2 + \beta^2 \right ]\,,
\end{equation}
crystallites of size $L$ produce a diffraction peak
\begin{equation}\label{eq:Ic}
    I_c(L,2\theta) = K V_c(L) A^{-1}(\beta)\mathcal{L}(2\theta)
\end{equation}
where $K$ contains all terms that are independent of the crystallite size for reflections of Bragg angle $\theta_B$. It provides diffraction peaks of fwhm $\beta$ and integrated intensity $P_c = \int I_c(L,2\theta)d2\theta =K V_c(L)$ since $\int \mathcal{L}(2\theta)d2\theta = A(\beta)$. For the Lorentzian line profile function in Eq.~(\ref{eq:lpf}),  $A(\beta)=\pi\beta/2$. As an example, $V_c(L) = L^3$ for crystallite with cubic shape of edge $L$, and the fwhm $\beta=0.92\lambda/L\cos\theta_B$ follows from  Eq.~(\ref{eq:SE}). Diffraction peaks from the powder samples are then given by
\begin{equation}\label{eq:I}
    I(2\theta) = \int I_c(L,2\theta)n(L)dL\,,
\end{equation}
and $P = \int I(2\theta)d2\theta = K\int V_c(L)n(L)dL = KV$ is proportional to the total volume $V$ of diffracting particles.

Experimental diffraction peaks from powder samples with PSD have widths $\beta_s$ defined by\cite{av19}
\begin{equation}\label{eq:Ihalf}
    I(2\theta_B\pm\beta_s/2)=\frac{1}{2}\int I_c(L,2\theta_B)n(L)dL = \int_0^{L_s}I_c(L,2\theta_B)n(L)dL
\end{equation}
where the median value $L_s$ is related to the fwhm $\beta_s$ through the SE. In other words, the experimental peak widths lead to the median value $L_s$ of the peak intensity weighted PSD, that is $I_c(L,2\theta_B)n(L)\propto L^4 n(L)$. Therefore, SE provides a measure of the median value of the weighted PSD by crystallite size to the power of four.\cite{av19} Before numerical demonstration that $\beta_s$ and $L_s$ in Eq.~(\ref{eq:Ihalf}) are connected by the SE, a brief discussion on how dynamical diffraction effects can be taken into account is given below.

\subsection{Dynamical diffraction effects}

\begin{figure}[ht]
  \includegraphics[width=3.2in]{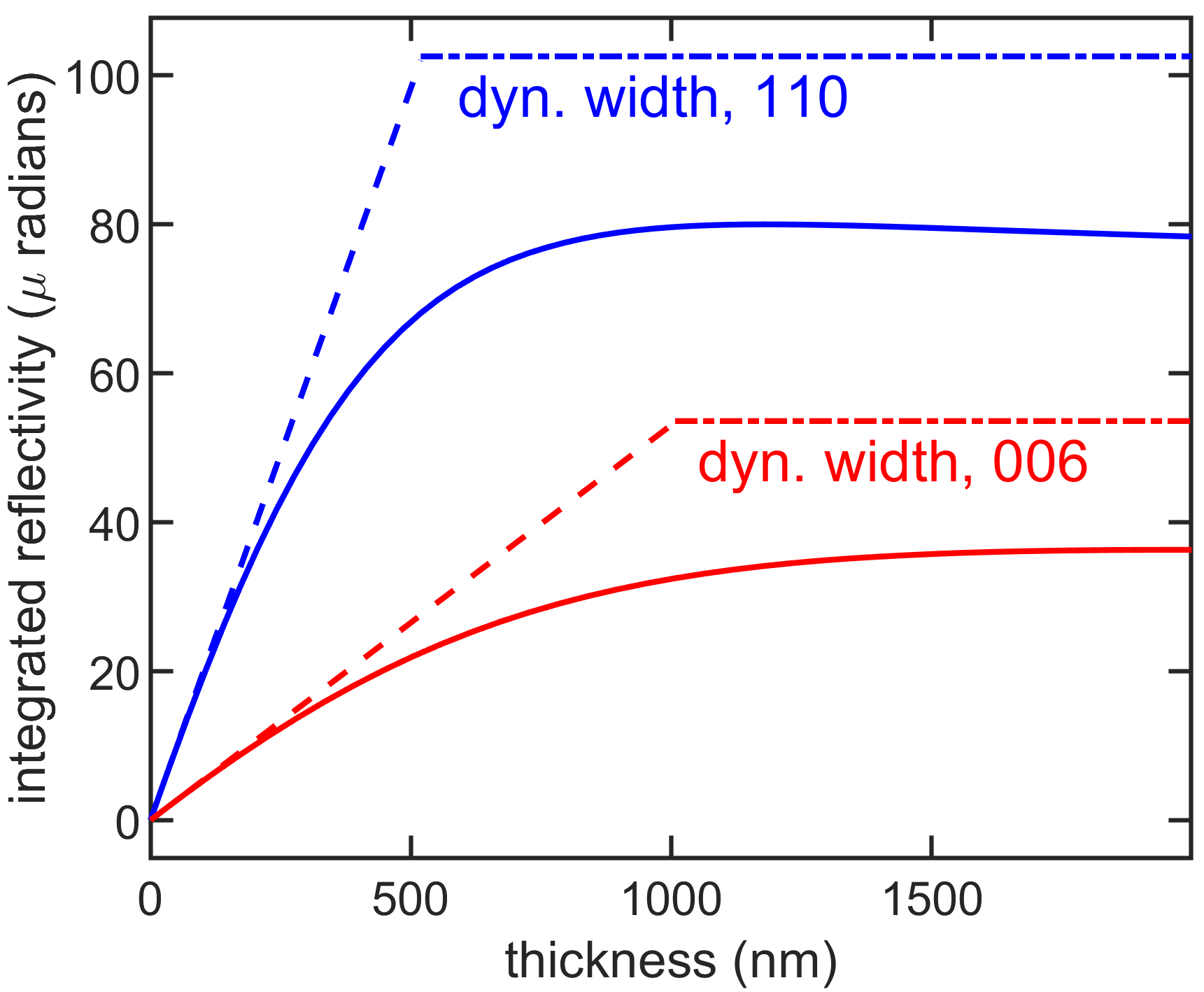}
  \caption{Integrated reflectivities (solid lines) from dynamical diffraction calculation as a function of thickness in \ce{BiFeO3} crystal slabs. X-rays of 12\,keV. For Bragg reflections 110 and 006, deviations of 5\% from the linear behaviour (dashed lines) due to dynamical diffraction effects (absorption and re-scattering processes) occur for slabs of thicknesses 132\,nm and 201\,nm, respectively. Maximum values of integrated reflectivities are always smaller than the intrinsic widths (dot-dashed lines) of the reflectivity curves.}
  \label{fig:kinlim}
\end{figure}

In a crystal slab of thickness $L^{\prime}$, the integrated reflectivity $P_{\rm dyn}(L^{\prime})$ from dynamical diffraction calculation in specular reflection geometry is always smaller than a finite value and it is proportional to $L^{\prime}$ only widthin the kinematical approach for very small crystals.\footnote[0]{Integrated reflectivities as a function of slab thickness can also be calculated via recursive series described elsewhere.\cite{sm17}} In more precise words,\cite{av19}
$$\lim\limits_{L^{\prime}\to\infty}P_{\rm dyn}(L^{\prime}) < W \quad{\rm and}\quad \lim\limits_{L^{\prime}\to0}P_{\rm dyn}(L^{\prime})=\alpha L^{\prime}$$
where $W$ is the intrinsic width\footnote[1]{In semi-infinite crystals (infinite thickness), integrated reflectivities have maximum values smaller than the intrinsic width $W$ of each Bragg reflection. It follows from the fact that reflectivity curves are always limited to values smaller than 1 so that their area $A<1\times W$.}$^,$\footnote[2]{$W=(r_e\lambda^2/V_{\rm cell})|F_{\rm hkl}F_{\bar{\rm h}\bar{\rm k}\bar{\rm l}}|^{1/2}/\sin2\theta_B$ for $\sigma$-polarized x-rays where $r_e=2.818\times10^{-5}$\,\AA\, is the classic electron radius, $F_{\rm hkl}$ is the structure factor of the hkl reflection with Bragg angle $\theta_B$ for the wavelength $\lambda$, and $V_{\rm cell}$ is the unit cell volume.\cite{sm16}} of a Bragg reflection and $\alpha$ is just a constant of proporcionality. Examples of dynamical integrated reflectivities, $P_{\rm dyn}(L^{\prime})$, are shown in Figure~\ref{fig:kinlim} for \ce{BiFeO3} (BFO) crystal with rhombohedral structure, space group $R3c$, and lattice parameters $a=b=5.57882$\AA, and $c= 13.8693$\AA. BFO is an important multiferroic material and nonlinear optical crystal with potential applications as second-harmonic nanoprobes for bio-imaging techniques.\cite{gc18}

To account for dynamical diffraction effects, the diffraction peak expression for crystallites of size $L$ in Eq.~(\ref{eq:Ic}) can be multiplied by the ratio $P_{\rm dyn}(L^{\prime})/\alpha L^{\prime}$, having in mind that $L^{\prime}$ stands for the crystallite dimension along the normal direction of the Bragg planes. In the case of cubic crystallites of edge $L$ and Bragg planes parallel to one face of the crystallites, dynamical diffraction effects are easily taking into account by rewriting Eq.~(\ref{eq:Ic}) as
\begin{equation}\label{eq:Icdyn}
    I_c(L,2\theta) = K \alpha^{-1}P_{\rm dyn}(L) L^2 A^{-1}(\beta)\mathcal{L}(2\theta)\,,
\end{equation}
in agreement with the kinematical approach where $\lim\limits_{L\to 0}\int I_c(L,2\theta)d2\theta = KV_c(L)$.

\subsection{Lognormal PSD}

The correlation between $\beta_s$ and $L_s$ in Eq.~(\ref{eq:Ihalf}) through the SE has been demonstrated in the case of a lognormal PSD,\cite{av19}
\begin{equation}\label{eq:logn}
    n(L) = \frac{N}{L\sigma\sqrt{2\pi}}\exp\left[-\dfrac{(\ln L - \ln L_h)^2}{2\sigma^2}\right]\,,
\end{equation}
often used to describe size distribution in systems of nanoparticles.\cite{lk99,ac05} $L_h=L_0\exp(\sigma^2)$ is the median value, that is $\int_0^{L_h}n(L)dL=N/2$, given in terms of both PSD parameters, the most probable particle size $L_0$ (mode) and $\sigma$ (the standard deviation in log scale). It follows from Eq.~(\ref{eq:Ihalf}) that, in the case of narrow PSDs where $\int I_c(L,2\theta_B)n(L)dL\approx I_c(L_0,2\theta_B)\int n(L)dL$, measures of diffraction peak widths in powder samples provide the particle size $L_s\simeq L_h$. However, for broader PSDs the particle size $L_s$ obtained from Scherrer equation has a more complex correlation with the parameters $L_0$ and $\sigma$, and where corrections for dynamical diffraction effects can be necessary.

\begin{figure*}[ht]
  \includegraphics[width=6.2in]{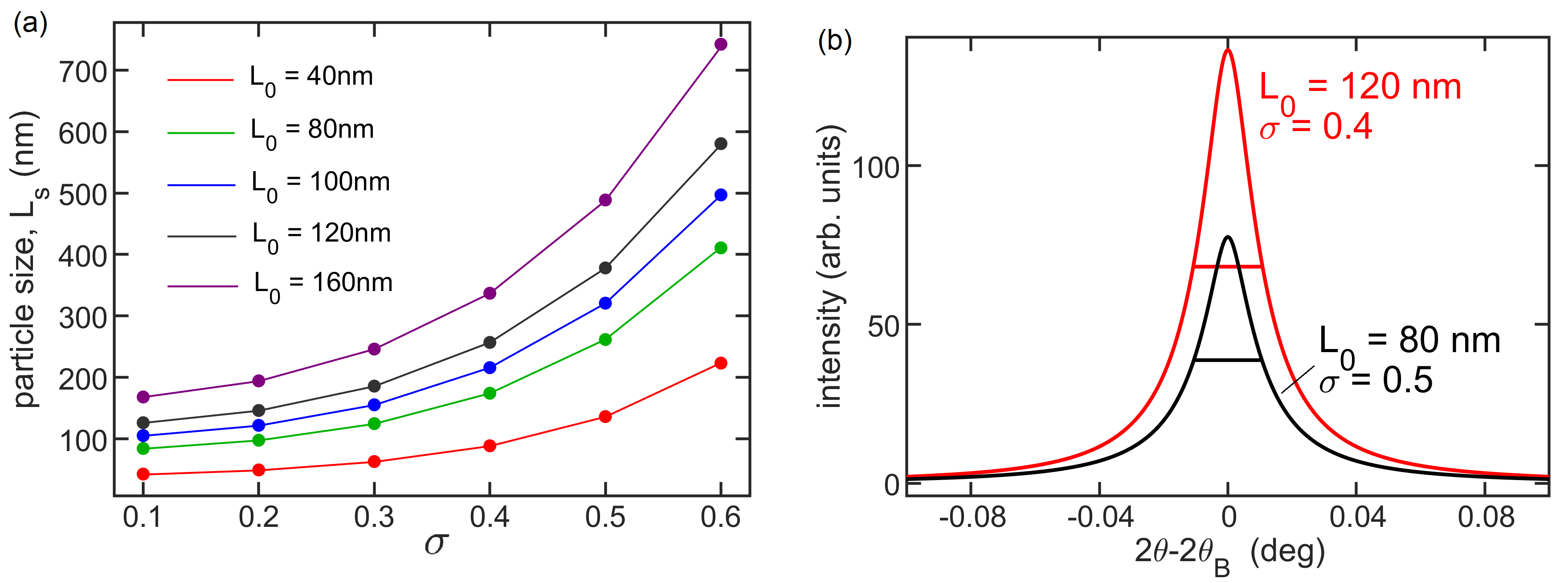}\\
  \caption{(a) Particle size (dots) from simulated peak width $\beta_s$ in Eq.~\ref{eq:I}, and median values $L_s$ (solid lines) from peak intensity weighted PSDs in Eq.~(\ref{eq:Ihalf}) as a function of lognormal PSD parameters $\sigma$ and $L_0$. (b) Simulated diffraction peaks $I(2\theta)$ of similar widths (horizontal lines) for PSDs with different parameters $L_0$ and $\sigma$, as indicated. Reflection 110 of the BFO crystal with X-rays of 12\,keV.}
  \label{fig:medianvalues}
\end{figure*}

\section{RESULTS AND DISCUSSION}

Measures of X-ray diffraction peak widths $\beta_s$ provide through the Scherrer equation $L_s = 0.92\lambda/\beta_s\cos\theta_B$ (for cubic crystallite), corresponding exactly to the median values of the peak intensity weighted PSDs, as defined in Eq.~(\ref{eq:Ihalf}). This correlation is numericaly demontrated in Figure~\ref{fig:medianvalues}a, taking as a reference the 110 reflection of the BFO crystal. Diffraction peaks $I(2\theta)$ are simulated from Eq.~(\ref{eq:I}) for Lorentzian line profile functions, Eq.~(\ref{eq:lpf}), and cubic crystallites with dynamical corrections as in Eq.~(\ref{eq:Icdyn}).

\begin{figure}[ht]
  \includegraphics[width=3.1in]{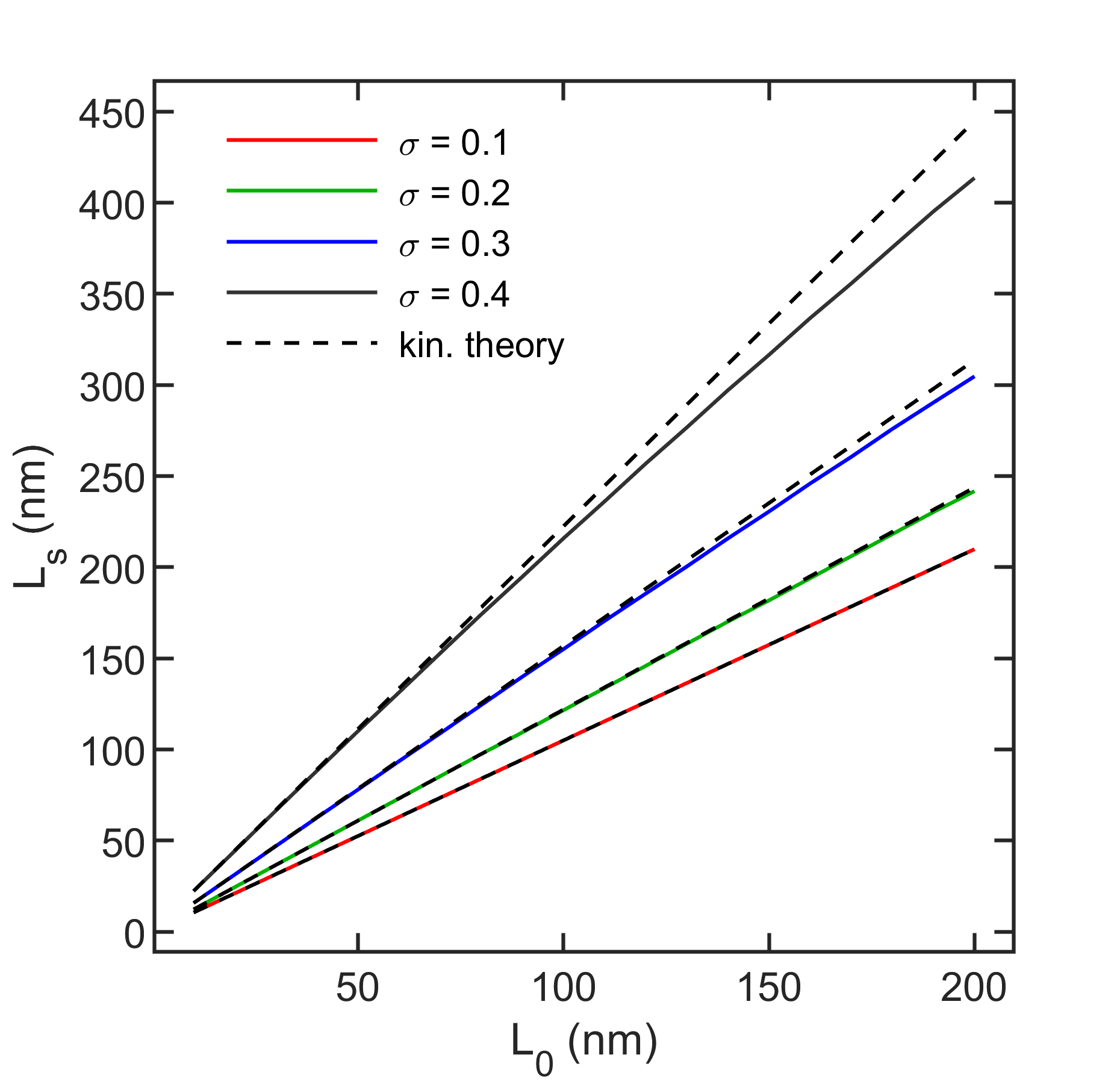} \\
  \caption{Particle size $L_s$ as a function of the most probable size $L_0$ in lognormal PSDs with different $\sigma$. Dynamical diffraction effects are disregarded in the kinematical approach (dashed lines). Calculation for 110 BFO reflection with x-rays of 12\,keV. }
  \label{fig:LsxSigma}
\end{figure}

Diffraction peak widths are unable to solve both parameters of the PSDs, for a single median value $L_s$ there are countless combinations of $L_0$ and $\sigma$. Examples of two simulated diffraction peaks of similar widths are shown in Figure~\ref{fig:medianvalues}b. When size distribution during crystallization can be described by nearly constant values of $\sigma$, it is possible to determine the PSD temporal evolution from the experimental peak widths by a nearly linear relationship as shown in Figure~\ref{fig:LsxSigma} for a few values of $\sigma$. In systems of BFO nanoparticles where diffraction peak widths led to $L_s\lesssim 120$\,nm, PSDs' mode can be given by $L_0 = \texttt{f}(\sigma) L_s$. For instance, $\texttt{f}(\sigma) = 0.954$, 0.824, and 0.645 for $\sigma=0.1$, 0.2, and 0.3, respectively. Although, integrated intensity values can undergo a reduction as large as 5\% for particles in this size range due to dynamical diffraction effects, Figure~\ref{fig:kinlim}, these effects demand no corrections in the $\texttt{f}(\sigma)$ coefficients as also shown in Figure~\ref{fig:LsxSigma}.

\section{CONCLUSIONS}

Peak width appearing in Scherrer's formula stands for median values of intensity-weighted size distributions (intensity = peak maximum intensity), that is the size distribution weighted by particles' dimension to the power of four. It implies that all current approaches where particles' morphology are estimated on bases of volume-weighted particle size distributions, that is weighting by particle's size to the power of 3, have to be corrected for weighting by particle's size to the power of 4. Establishing a direct correlation between peak width and size distribution allows tracking temporal evolution of size and size distribution during in situ studies of crystallization processes, making it conceivable to distinguish periods of nucleation, coarsening, and Ostwald ripening. Corrections for dynamical effects can be neglected for systems of nanoparticles with sizes below 100\,nm, even when the integrated intensities diminish during in situ studies as recently observed.\cite{ac19}

\bibliography{ScherrerFormulaPSD}

\end{document}